\documentclass[preprint,sort&compress,3p,onecolumn]{elsarticle}
\usepackage{amssymb,amsbsy,amsmath,amsfonts,multirow}
\usepackage{graphicx}
\usepackage{dcolumn}
\usepackage{bm}
\usepackage{color}

\usepackage{hyperref}
\hypersetup{
  colorlinks=true,        
  linkcolor=blue,         
  citecolor=cyan,         
  urlcolor=red,           
}

\begin{document}

\begin{frontmatter}

\title{On the Existence of $N^*(890)$ Resonance in $S_{11}$ Channel of $\pi N$ Scatterings}

\author[Wang]{Yu-Fei Wang}
\address[Wang]{Department of Physics and State Key Laboratory of Nuclear Physics and Technology, \\
Peking University, Beijing 100871, China}

\author[Yao]{De-Liang Yao}
\address[Yao]{Instituto de F\'{\i}sica Corpuscular (centro mixto CSIC-UV), \\
Institutos de Investigaci\'{o}n de Paterna, \\
Apartado 22085, 46071, Valencia, Spain}
\author[Wang,Zheng]{Han-Qing Zheng}
\address[Zheng]{Collaborative Innovation Center of Quantum Matter, \\
Beijing 100871, China}

\begin{abstract}
Low-energy partial-wave $\pi N$ scattering data is reexamined with the help of the production representation of partial-wave $S$ matrix, where branch cuts and poles are thoroughly under consideration. The left-hand cut contribution to the phase shift is determined, with controlled systematic error estimates, by using the results of $\mathcal{O}(p^3)$ chiral perturbative amplitudes obtained in the extended-on-mass-shell scheme. In $S_{11}$ and $P_{11}$ channels, severe discrepancies are observed between the phase shift data and the sum of all known contributions. Statistically satisfactory fits to the data can only be achieved by adding extra poles in the two channels. We find that a $S_{11}$ resonance pole locates at $\sqrt{z_{r}}=(0.895\pm0.081)-(0.164\pm0.023)i$ GeV, on the complex $s$-plane. On the other hand, a $P_{11}$ virtual pole, as an accompanying partner of the nucleon bound-state pole, locates at $\sqrt{z_{v}}=(0.966\pm0.018)$ GeV, slightly above the nucleon pole on the real axis below threshold. Physical origin of the two newly established poles is explored to the best of our knowledge.  It is emphasized  that the $\mathcal{O}(p^3)$ calculation greatly improves the fit quality comparing with the previous $\mathcal{O}(p^2)$ one.
\end{abstract}


\end{frontmatter}

The studies in the field of $\pi N$ scatterings have led to and  fertilized many useful ideas and methods in the development of hadron and even particle physics, such as the establishment of dispersion techniques~\cite{Chew:1957zz}, the concepts of finite energy sum rules~\cite{Logunov:1967dy,Igi:1967zza} and duality~\cite{Dolen:1967jr}. In decades, a wealth of experimental data on differential cross section and polarizations~\cite{Hohler:1983} has been accumulated. Partial wave analysis of the $\pi N$ scattering amplitudes has also been performed which results in the discovery of many baryon resonances~\cite{Koch:1980ay,Koch:1985bn,Matsinos:2006sw,Arndt:2006bf}.

Efforts in modern studies on $\pi N$ scatterings focus on a precision description of the $\pi N$ amplitudes, not only in the physical region but also in the subthreshold region. Baryon chiral perturbation theory (BChPT)~\cite{Jenkins:1990jv,Bernard:1995dp,Bernard:2007zu} serves as an appropriate tool for such a purpose, at low energies. The problem of power counting breaking terms~\cite{Gasser:1987rb} in loops, caused by the presence of baryon degrees of freedom in the chiral Lagrangian, is finally overcome  by the use of the extended-on-mass-shell subtraction scheme~\cite{Fuchs:2003qc}. Though many achievements have been obtained in the perturbation calculations within this scheme~\cite{Alarcon:2012kn,Chen:2012nx,Yao:2016vbz,Siemens:2016hdi,Siemens:2017opr}, analyses respecting both exact partial wave unitarity and proper analyticity are still lacking.  Hence the use of the dispersion techniques revived in recent years. For example, the analysis based on Roy-Steiner equations~\cite{Hoferichter:2015hva}  has reproduced low-energy phase shift data successfully (see also Refs.~\cite{Gasparyan:2010xz,Mathieu:2015gxa}).

More recently, in Ref.~\cite{Wang:2017agd},  low-energy $\pi N$ scattering has been re-visited with the help of the Peking University (PKU) representation~\cite{Xiao:2000kx,Zheng:2003rw,Zhou:2006wm,Zhou:2004ms} -- a production parametrization of the unitary partial wave $S$ matrix  for two-body elastic scattering amplitudes. The production representation is constructed from first principles. Especially, analyticity and unitarity are automatically built in, hence the fatal deficiency of the violation of analyticity in more often used unitarization approximation method is avoided~\footnote{The PKU representation has been successfully used to settle down the long standing issue on the existence of  $f_0(500)$ meson~\cite{Xiao:2000kx,Zhou:2006wm} and  $\kappa(800)$ meson~\cite{Zheng:2003rw,Zhou:2004ms}, by pointing out that the $f_0(500)$ meson is essential to adjust chiral perturbation theory to experiments~\cite{Xiao:2000kx}. It predicts the pole locations in almost perfect coincidence with results from the Roy equation analysis~\cite{Caprini:2005zr,DescotesGenon:2006uk}.}. In addition, the representation is fully compatible with crossing symmetry~\cite{Guo:2007ff,Guo:2007hm}.

The advantage of PKU representation is apparent. It separates partial waves into various terms contributing either from poles or branch cuts. The consequent phase shifts are sensitive to the location of subthreshold poles, enabling one to determine pole positions rather accurately. Furthermore, each phase shift contribution has a definite sign, which makes possible the disentanglement of hidden poles from a background~\footnote{The hidden poles stand for the $S$-matrix singularities, such as virtual states, resonances below the threshold or with  large widths, or shadow pole structures, which can hardly be observed from other approaches. }. In PKU representation, the background cut contribution is estimated from perturbative BChPT amplitudes and uncertainties from such an estimation is known to be suppressed. It is important to emphasize that the philosophy is not to directly unitarize the amplitude itself, instead, it unitarizes the left-hand cuts of the chiral perturbative amplitudes. Hence hazardous spurious poles, which violates causality and spoils the determination of subthreshold poles, are absent.

\begin{figure*}[t]
\includegraphics[width=0.45\textwidth]{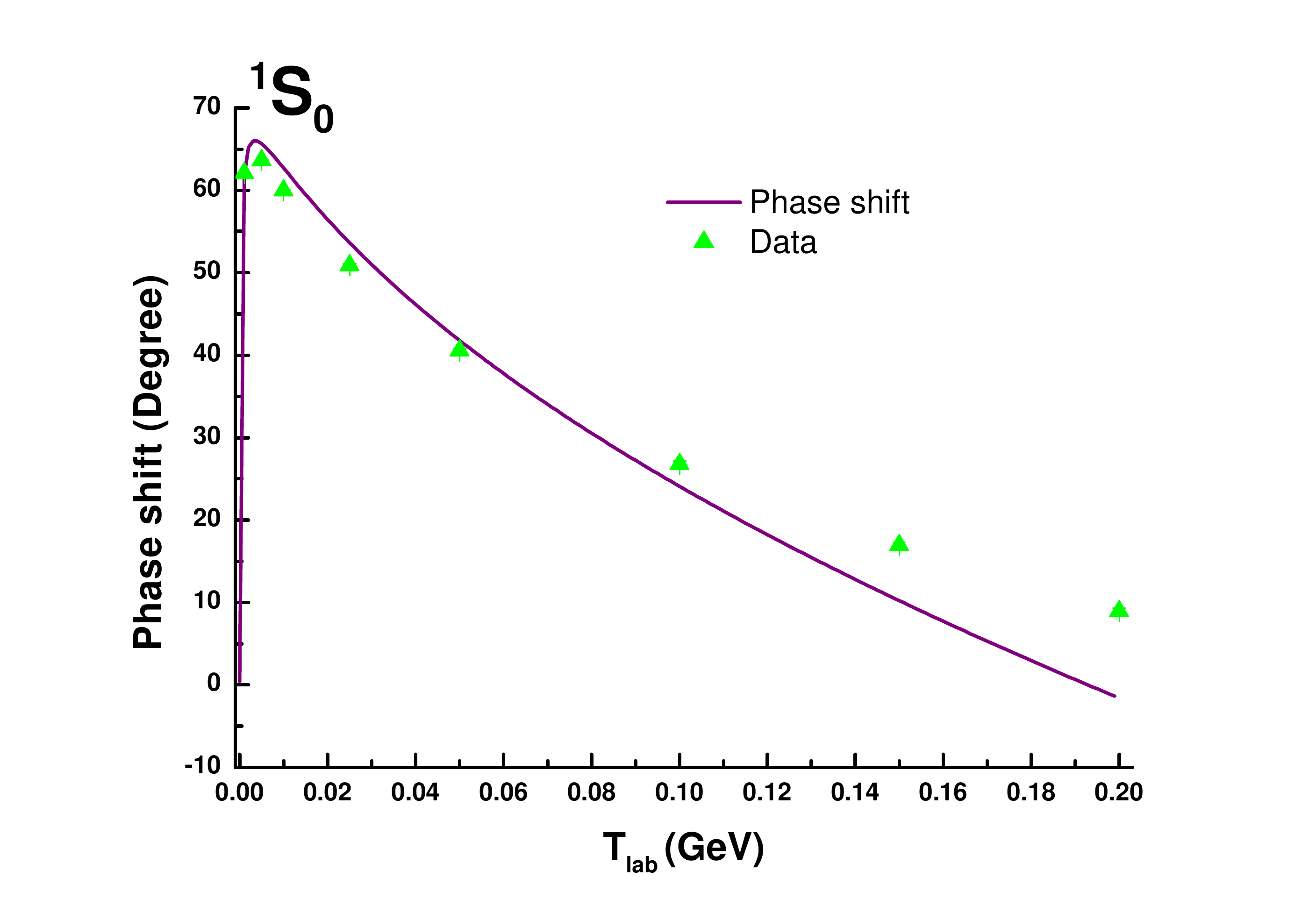}
\includegraphics[width=0.45\textwidth]{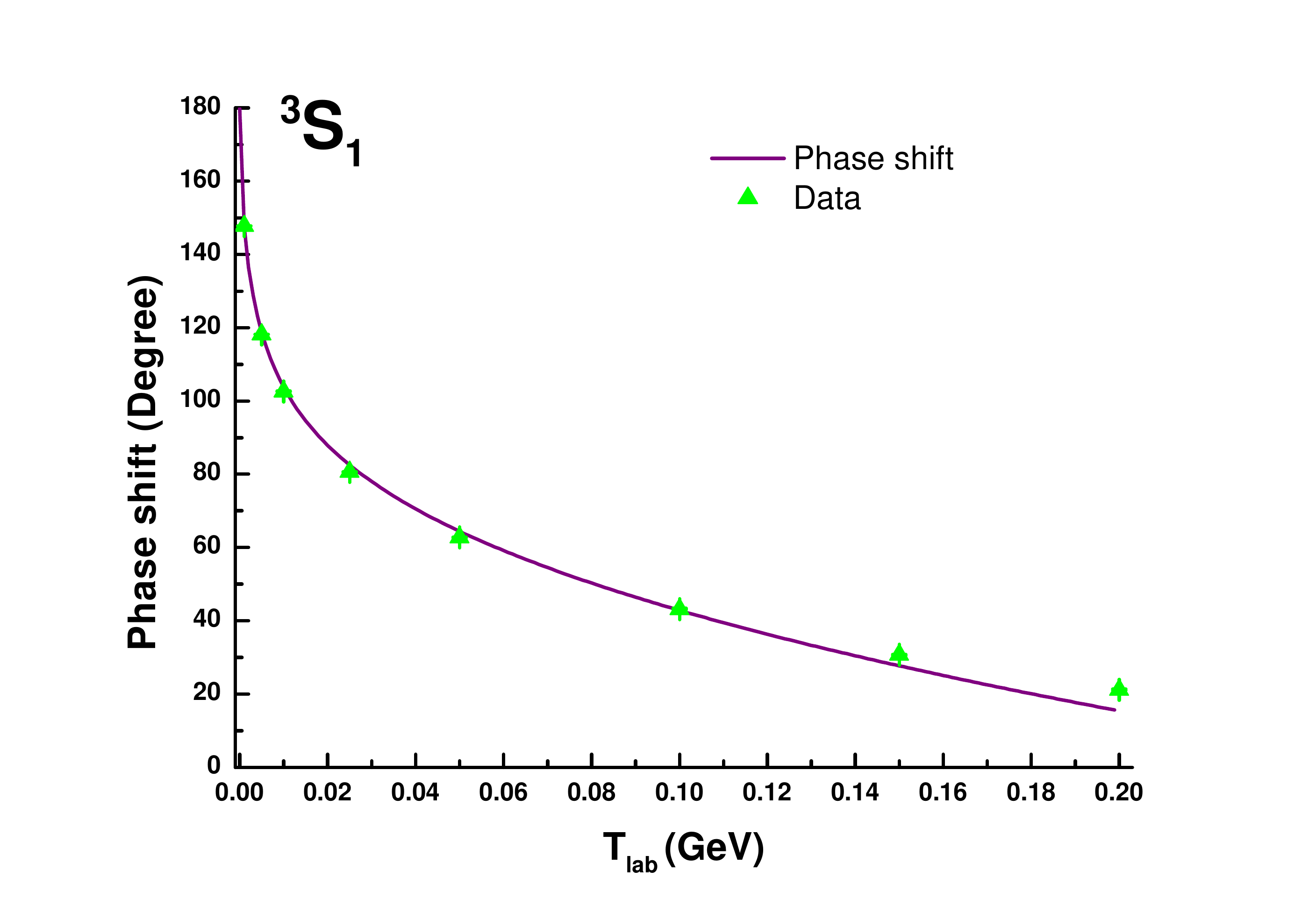}
\caption{\label{fig:hufit} The fit to $NN$ phase shift data with Ning Hu representation. }
\end{figure*}

The application of PKU representation is not confined to meson--meson~\cite{Xiao:2000kx,Zheng:2003rw} or meson--baryon interactions~\cite{Wang:2017agd} only. In the classical field of nucleon--nucleon scatterings, an illuminating example can clearly re-depict the major physical picture therein. For $^1S_0$ and $^3S_1$ channels the following $S$-matrices are imposed respectively~\footnote{This is the Ning Hu representation in quantum mechanical scattering theory, which can be viewed as a simplified version of the PKU representation. See N. Hu, Phys. Rev. {\bf 74}, 131 (1948); T. Regge, Nuovo Cim. {\bf 8}, 671 (1958). See also, Pages~363-366 of R. G. Newton, {\it Scattering Theory of Waves and Particles}, second edition, Springer-Verlag, New York Heidelberg Berlin, 1982.
}
\begin{eqnarray}\label{eq:Hu}
S(^1S_0)=\frac{iv-k}{iv+k}e^{-2iR_0 k}\ ,\,\,\,S(^3S_1)=\frac{id+k}{id-k}e^{-2iR_{1} k}
\end{eqnarray}
where $-iv$ and $id$ are the pole positions of the $^1S_0$ virtual state and the deuteron on the $k$ plane (with $v,d>0$), and $R_0$, $R_{1}$ are two parameters (background constants) that are of the order of typical range of nuclear force, i.e. $R_{0,1}\sim 1$ fm. The binding energies in terms of $v$ and $d$ are $B_v=-v^2/M_N$ and $B_d=-d^2/M_N$ ($M_N$ is the nucleon mass). Fit results to the data of Ref.~\cite{Stoks:1993tb} are shown in Fig.~\ref{fig:hufit}.
Despite being equipped with such a compact parametrization,  the data is still qualitatively described with the following reasonable outcomes:
$R_0=1.01\pm0.05\ \text{fm},\ R_{1}=0.927\pm0.026\ \text{fm},\ v=(8.66\pm0.31)\ \mbox{MeV, }d=\ (43.1\pm0.64)\ \mbox{MeV}. $
The positions of $^1S_0$ virtual state and the deuteron pole are in agreement with the values given in Refs.~\cite{Stoks:1993tb,Entem:2016ipb} within  $10\%$. The fit also leads to the scattering lengths ($a$) and effective ranges ($r$):
$a_0=-21.74\pm0.86\ \text{fm},\ r_0=2.12\pm0.10\ \text{fm},\ a_{1}=5.49\pm0.04\ \text{fm},\ r_{1}=1.56\pm0.03\ \text{fm}. $

The above example demonstrates the powerfulness of the PKU representation in revealing subthreshold or low-lying distant poles. Returning to the study of $\pi N$ scatterings, in Ref.~\cite{Wang:2017agd}
the application of the PKU representation has led to very interesting findings: a resonance below $\pi N$ threshold in the $S_{11}$ channel is called for to saturate the huge discrepancy between the phase shift data and the known contributions, and a companionate virtual state of the nucleon in the $P_{11}$ channel is found.

However, the conclusions are drawn in Ref.~\cite{Wang:2017agd} with the left-hand cuts estimated by tree-level BChPT amplitudes. That is, only the kinematical cut $(-\infty,(M_N-m_\pi)^2]$ and the segment cut $[(M_N^2-m_\pi^2)^2/M_N^2,M_N^2+2m_\pi^2]$ due to the $u$-channel nucleon exchange are taken into account, where $M_N$ and $m_\pi$ are physical masses of the nucleon and the pion, respectively. As a result it is difficult to achieve a good $\chi^2$ fit for all the six $S$- and $P$-wave channels. Hence, to be serious, the observation in Ref.~\cite{Wang:2017agd} is better to be considered as a suggestion rather than a robust conclusion. Here we go beyond the work of Ref.~\cite{Wang:2017agd} by exploring the full structure of the left-hand cuts for $\pi N$ scatterings at ${\cal O}(p^3)$ level.  The implementation of the full structure of the left-hand cuts affords further and crucial evidence in support of the very existence of the two hidden poles in the $S_{11}$ and $P_{11}$ channels. A novel resonant state $N^*(890)$ with $J^P=\frac{1}{2}^-$ in $S_{11}$ channel can be claimed to be established.

The PKU representation of the partial-wave $\pi N$ elastic scattering $S$ matrix can be written as
\begin{equation}\label{eq:Spku}
S(s)=\prod_bS_b(s)\cdot\prod_vS_v(s)\cdot\prod_rS_r(s)\cdot e^{2\text{i}\rho(s)f(s)}\ \mbox{, }
\end{equation}
with $\rho(s)={\sqrt{s-s_L}\sqrt{s-s_R}}/{s}$, and $s_L=(M_N-m_\pi)^2$, $s_R=(M_N+m_\pi)^2$. The indices $b$, $v$ and $r$ denote bound states, virtual states and resonances, respectively (see Ref.~\cite{Wang:2017agd} for detailed expressions).
The exponential term is a background that carries the information of left-hand cuts (\textit{l.h.c.}s) and right-hand inelastic cut {(\textit{r.h.i.c.})} above inelastic thresholds. The function $f(s)$ satisfies a dispersion relation,

\begin{eqnarray}
f(s)&&=-\frac{s}{\pi}\int_{s_{c}}^{(M_N-m_\pi)^2} \frac{\ln|S(s')|ds'}{2\rho(s')s'(s'-s)}
+\frac{s}{\pi}\int_0^{\theta_c}\frac{\ln[S_{in}(\theta)/S_{out}(\theta)]}{2i\rho(s')(s'-s)\mid_{s'=(M_N^2-m_\pi^2)e^{i\theta}}}d\theta\nonumber\\
&&+\frac{s}{\pi}\int_{(M_N^2-m_\pi^2)^2/M_N^2}^{M_N^2+2m_\pi^2} \frac{\text{Arg}[S(s')]ds'}{2is'\rho(s')(s'-s)}
+\frac{s}{\pi}\int_{(2m_\pi+M_N)^2}^{\Lambda_R^2} \frac{\ln[1/\eta(s')]ds'}{2\rho(s')s'(s'-s)}\ \mbox{, }\label{fdisperpiNp3}
\end{eqnarray}

where $S_{in}$ and $S_{out}$ are the $S$-matrix values calculated from BChPT amplitudes inside and outside along the circular cut, respectively; and $0<\eta<1$ is the inelasticity along the \textit{r.h.i.c.} starting from $(2m_\pi+M_N)^2$. The $s_c$ and $\theta_c$ are two cut-off parameters for the \textit{l.h.c.}s, while $\Lambda_R$ is a cut-off parameter for the \textit{r.h.i.c.}.

At $\mathcal{O}(p^3)$ level, the first term on the right-hand side of Eq.~(\ref{fdisperpiNp3}) is negative definite and dominates the $f(s)$ contribution. The second term (circular cut) is a new contribution compared with the study of Ref.~\cite{Wang:2017agd}. This term and the third term are found to be numerically small: the former appears purely at $\mathcal{O}(p^3)$ level, while the latter stems from $u$-channel nucleon exchange diagrams, which can be approximated as contact interaction in the low energy region. In contrast, the fourth term, coming from \textit{r.h.i.c.}, is positive definite, being able to be extracted from known data of phase shifts and inelasticities, but does not impact significantly at low energies. Eventually, the function $f(s)$ receives a negative net contribution to the phase shift with any reasonable choices of the cut-off parameters. In our case, two {\it ad hoc} values $s_c=-0.08$ GeV$^2$ and $\theta_c=1.18$ radian can be imposed according to the valid region where BChPT is assumed to work (up to the center-of-mass (CM) energy $W\equiv\sqrt{s}\leq 1.4$ GeV, which is a bit optimistic). For the details of how to determine contributions from cuts and known poles, see Refs.~\cite{Wang:2018,Wang:2017agd}.

The fact that $f(s)$ contributes negatively to the phase shift is actually of crucial importance in exposing hidden poles: the only other source of negative contribution is a bound state, which is seen experimentally and can hence be fixed in the present scheme. Consequently, hidden poles on the second Riemann sheet, giving positive definite contributions, can be readily distinguished from a negative background. We in the following discuss the physics of $P_{11}$ and $S_{11}$ waves separably.

\begin{figure*}[htbp]
\centering
\includegraphics[width=0.45\textwidth]{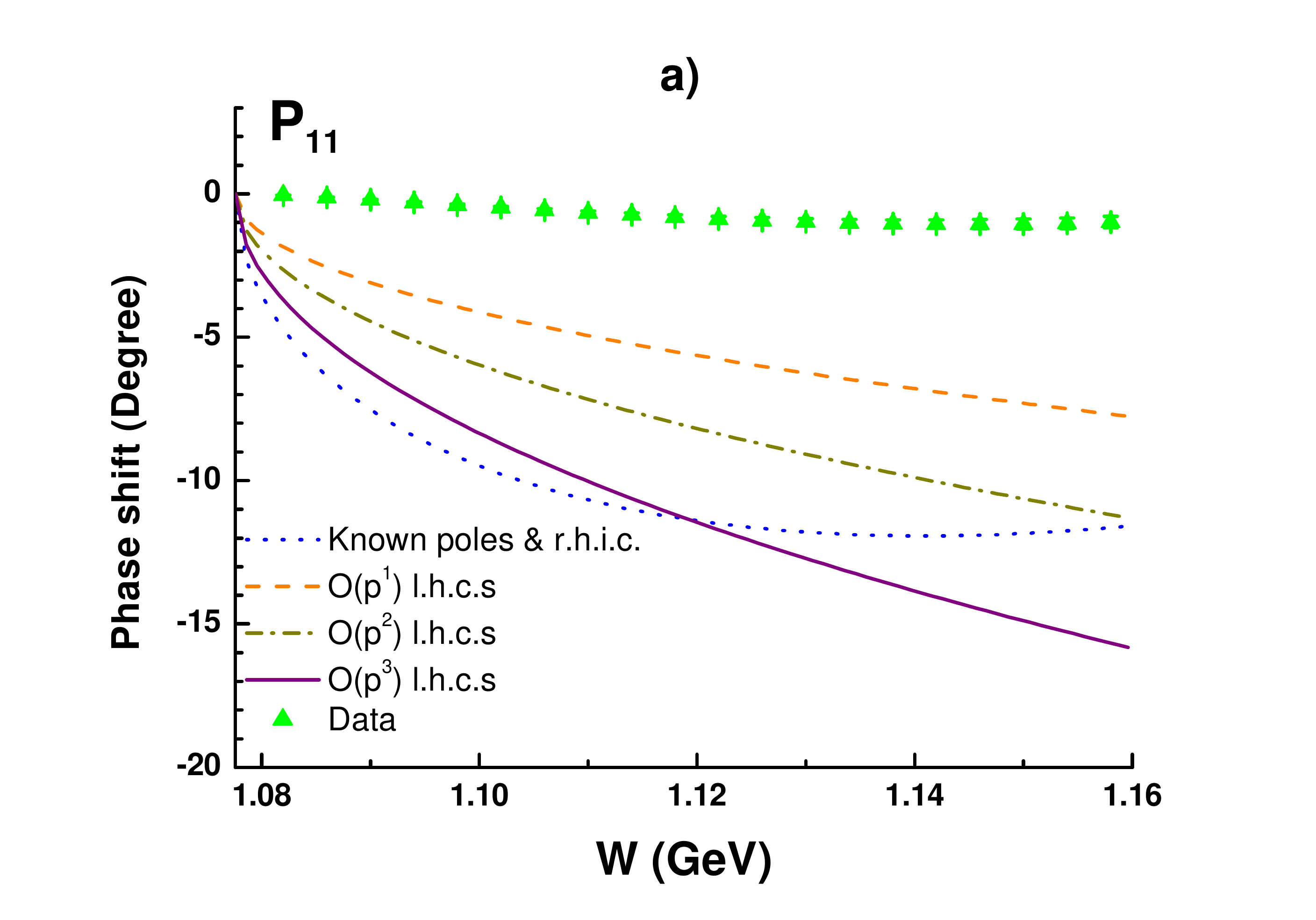}
\includegraphics[width=0.45\textwidth]{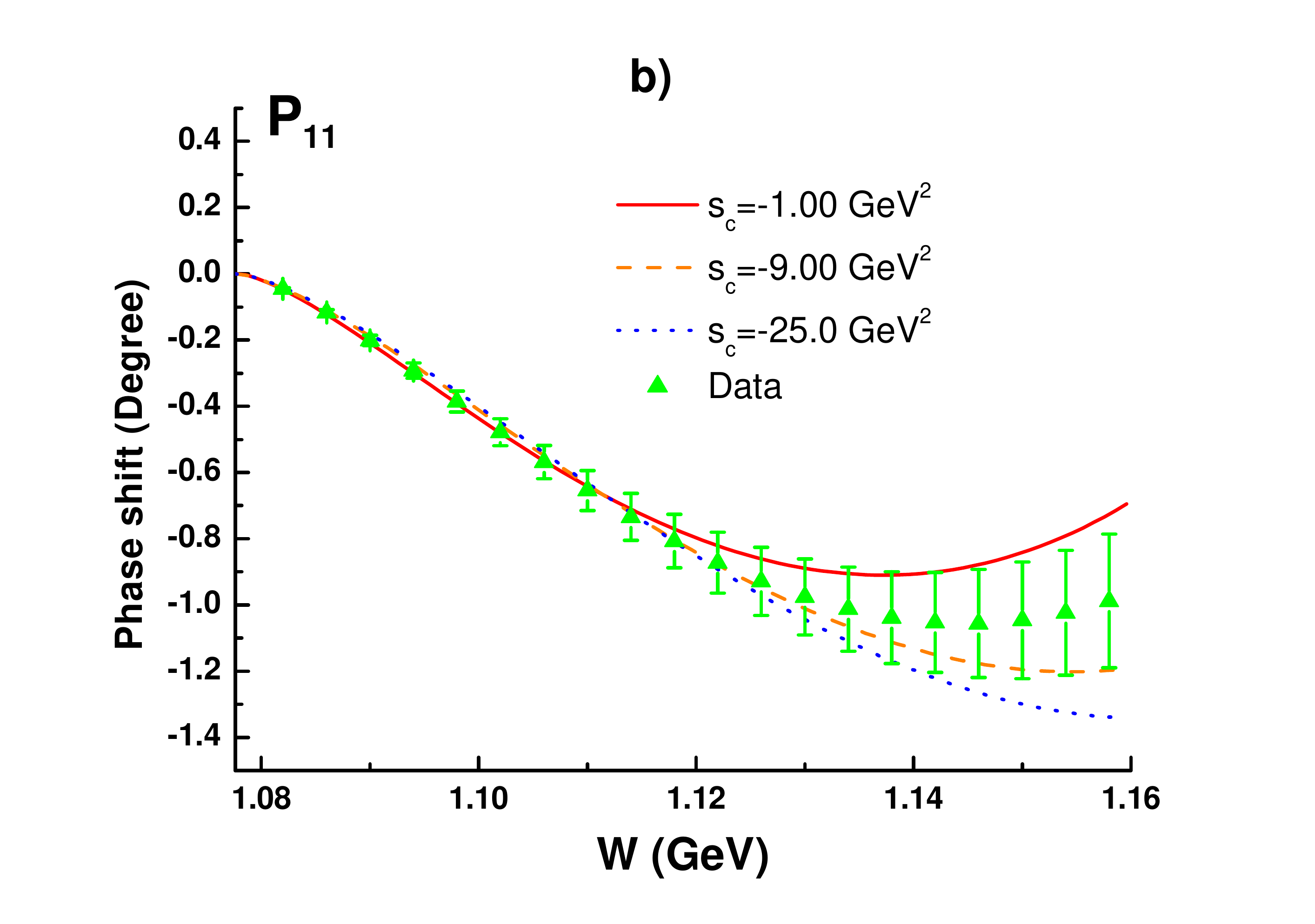}
\caption{\label{fig:P11ana} Different contributions in $P_{11}$ channel. a): known contributions vs data~\cite{Hoferichter:2015hva} ($s_c=-0.08$ GeV$^2$). b): fits to the data with the resonance under different choices of the cut-off parameter. }
\end{figure*}

{\it $P_{11}$ wave:} The phase shift is very small, i.e., $|\delta(s)|\leq 1$~degrees, for any CM energy $W$ in the range from $\pi N$ threshold to $1.16$~GeV, this $\delta(k)=\mathcal{O}(k^3)$ ($k$ being the $3$-momentum) behaviour has to come from cancelations of different contributions. In the following we demonstrate that the cancelations can only happen owing to the existence of a near-threshold virtual pole. In Fig.~\ref{fig:P11ana}-(a), various known-pole and cut contributions are displayed. In particular, the \textit{l.h.c.}s are shown with the BChPT inputs at different chiral orders. They are all sizeably negative, and obviously their sum is at odds with the existing data. Moreover, the deviation becomes more significant when the $\mathcal{O}(p^3)$ contribution is taken into consideration. Such a large deviation requires new structures in the $S$ matrix, contributing positively to the phase shift.  Hidden virtual poles meet this requirement. Similar to Ref.~\cite{Wang:2017agd}, two virtual states (or a resonance) are added to the $S$ matrix. By fitting to the phase shift data, one virtual pole stays in the near-threshold region as main hidden contribution, while the other is nearly absorbed by the pseudo-threshold $(M_N-m_\pi)^2$~\footnote{Even if we put a resonance, it automatically converts into two virtual states. This fact supports crucially the stability of our whole program. As already discussed in Ref.~\cite{Wang:2017agd}, one can also pin down the location of the virtual state by perturbation calculation or subthreshold expansion, and the two results coincide with each other, implying that such a way of estimating \textit{l.h.c.}s is fully justified. }. In Fig.~\ref{fig:P11ana}-(b), the fit results are shown. In the fits for each preferred $s_c$ value, a $P$-wave threshold constraint, i.e. $\delta(k)=\mathcal{O}(k^3)$, is considered. A good fit can only be obtained by decreasing the {\it ad hoc} value $s_c=-0.08$ GeV$^2$ slightly down to  $s_c=-1.00$ GeV$^2$. However, the $\mathcal{O}(p^2)$ calculation in Ref.~\cite{Wang:2017agd} needs a rather large $|s_c|$ like $s_c=-9.00$ GeV$^2$. In other words, the $\mathcal{O}(p^3)$ calculation certainly improves the fit quality in $P_{11}$ in the sense that the low-energy fit now no longer relies much on the uncertain high-energy contributions~\footnote{In principle the cut-off parameter $s_c$ can also be set as a fit parameter, but due to the existence of many local-minimums, such fit does not converge and the statistical errors are no longer trustable, though the physical outputs change little. }.  The dependence of the location of the near-threshold virtual state on the cut-off parameter is examined as well, see fits with different $s_c$ in Fig.~\ref{fig:P11ana}-(b), which is responsible for the systematical uncertainty. The statistical error from fits is negligible. Finally, the pole location is determined to be
\begin{equation}
\sqrt{z_{v}}=(0.966\pm0.018)~ {\rm GeV}\ .
\end{equation}
 The physical interpretation of this virtual state is already revealed in Ref.~\cite{Wang:2017agd} -- it can actually be understood as a kinematic companionate pole of the nucleon bound state.

\begin{figure*}[t]
\centering
\includegraphics[width=0.45\textwidth]{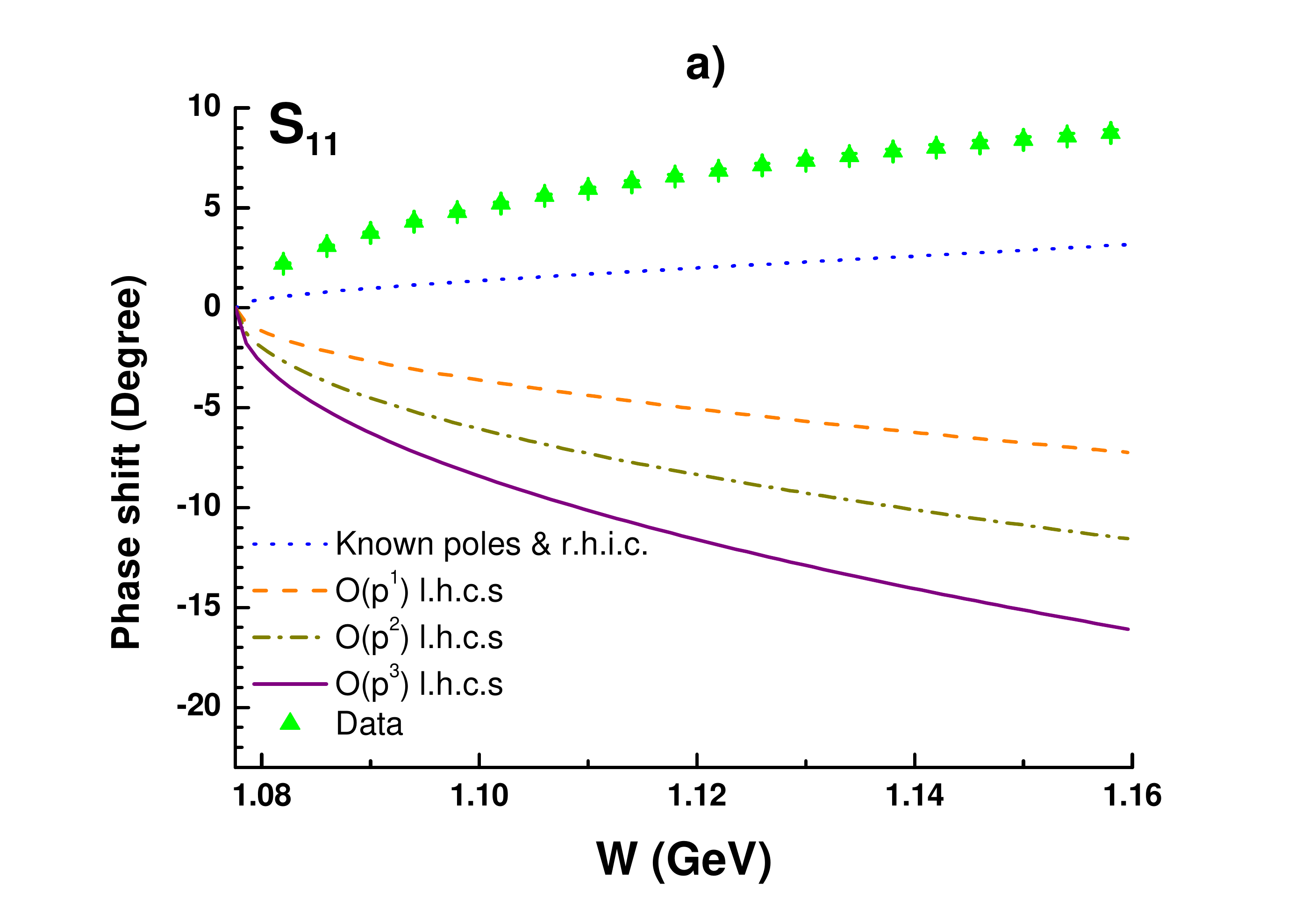}
\includegraphics[width=0.45\textwidth]{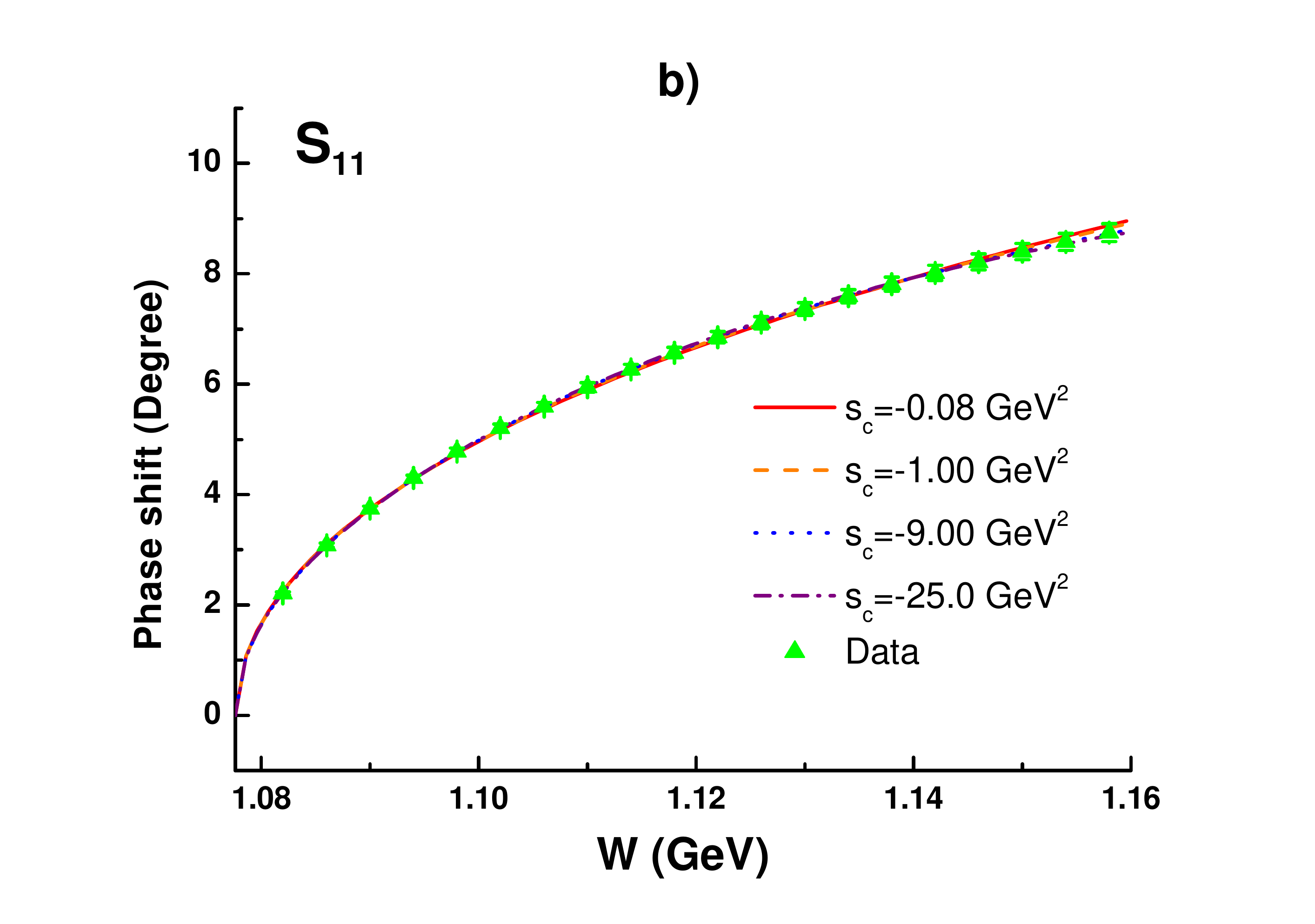}
\caption{\label{fig:S11ana} Different contributions in $S_{11}$ channel. a): known contributions vs data~\cite{Hoferichter:2015hva} ($s_c=-0.08$ GeV$^2$). b): fits to the data with the resonance under different choices of the cut-off parameter. }
\end{figure*}
{\it $S_{11}$ wave:}  The phase shift data show a rapid increase in the threshold region. It is abnormal since the nearest known resonance is $N^\ast(1535)$, far away form the $\pi N$ threshold and unlikely providing such a significant  positive growth in that region. As can be seen from  Fig.~\ref{fig:S11ana}-(a), all the known poles (including $N^\ast(1535)$) plus \textit{r.h.i.c.} indeed can not afford enough effects to achieve the above-mentioned rapid increase.  Further taking into account the \textit{l.h.c.}s, the $S_{11}$ channel suffers from a larger discrepancy between the known contributions and the data, which can be rescued neither by any fine-tuning in the numerical analyses, nor by promoting the BChPT inputs to higher chiral orders, as can be readily witnessed from Fig.~\ref{fig:S11ana}-(a). Nonetheless, the abnormal enhancement can be resolved by the introduction of a novel resonance in the $S_{11}$ channel on top of the other well-known excitations. The position of this hidden resonance is determined by the fit to data with different cut-off parameters, which persistently indicates the existence of a very stable ``crazy resonance'' below threshold with similar fit qualities, no matter how one changes the cut-off parameter,
as shown in Fig.~\ref{fig:S11ana}-(b). Our final result on this hidden resonance reads
\begin{equation}
\sqrt{z_{r}}=(0.895\pm0.081)-(0.164\pm0.023)i~ {\rm GeV}\ ,
\end{equation}
which is compatible with the one from the $\mathcal{O}(p^2)$ analysis~\cite{Wang:2017agd}, i.e. $(0.861\pm0.053)-(0.130\pm0.075)i$ GeV. We name this new resonance structure as $N^*(890)$.

Thus,  a thorough consideration of the cut structure at $\mathcal{O}(p^3)$ level corroborates the existence of the hidden poles in $S_{11}$ and $P_{11}$ channels, with proper systematical uncertainty estimates. Compared to the $\mathcal{O}(p^2)$ analysis in Ref.~\cite{Wang:2017agd}, the $\mathcal{O}(p^3)$ contribution plays an essential role in performing a meaningful numerical fit, from which trustworthy determination on the positions of the hidden poles can be obtained. Besides, it is also indispensable for other $S$- and $P$-wave channels: similarly to $P_{11}$ wave, it is found that $\mathcal{O}(p^2)$ does not fit well for low $s_c$ in $P_{31}$ channel, whereas $\mathcal{O}(p^3)$ does; in addition, in channels like $S_{31}$ and $P_{13}$, without $\mathcal{O}(p^3)$ contribution, the chi-squares would be unacceptably large. Only in $P_{33}$ channel the description of data is not very much relevant to chiral orders. It is also worth noticing here that in $S_{31}$ channel certain singularity structure may be needed in accommodating data. More details are referred to Ref.~\cite{Wang:2018}. As side products, for all the $S$- and $P$-waves we calculate the $R$ values, representing the interaction range of the $\pi N$ potential (c.f. Eq.~(\ref{eq:Hu})),  which are collected in Table~\ref{tab:HuR}.
The resulting $R$ values are of the order $1/(2m_\pi)$, which turn out to be reasonable since the two-pion exchange is the lowest contribution in the $t$-channel $\pi N$ interactions.

\begin{table}[t]%
\caption{\label{tab:HuR}%
$R$ parameters for $S$- and $P$-wave  $\pi N$ potentials.
}
\begin{center}
\begin{tabular}{ccccccc}
\hline
\textrm{Channel}&\textrm{$S_{11}$}&\textrm{$S_{31}$}&\textrm{$P_{11}$}&\textrm{$P_{31}$}&\textrm{$P_{13}$}&\textrm{$P_{33}$}\\
$R$  ($m_{\pi}^{-1}$) & $0.56$ & $0.47$ & $0.52$ & $0.38$ & $0.38$ & $0.30$\\
\hline
\end{tabular}
\end{center}
\end{table}

In the end, it should be stressed that though one may expect  an $\mathcal{O}(p^4)$ calculation could even further improve the fit quality, it is no longer in urgent need, since the fit quality at $\mathcal{O}(p^3)$ level is already satisfactory. Especially, no extra analytic structure is expected to show up in the $S$ matrix at $\mathcal{O}(p^4)$ level.

In summary, this letter provides a crucial evidence in support of the existence of the $P_{11}$ virtual pole, an accompanying partner of the nucleon bound state, and the subthreshold $S_{11}$ resonance $N^*(890)$. The $P_{11}$ virtual pole, associated with the ``elementary'' nature of the nucleon bound state, has been overlooked in decades of studies. For the novel $S_{11}$ resonance, it was suggested in Ref.~\cite{Wang:2017agd} that it might be a potential-like resonance. However, since it is too deep, a correct understanding of its nature remains open. The exsitence of the two hidden poles offers an natural explanation on the low-energy behaviour of the phase shifts in the $P_{11}$ and $S_{11}$ waves, and will hopefully open a new window for studies in the field of $\pi N$ phenomenology.

{\it Acknowledgments:} One of the authors (YFW) acknowledges Helmholtz-Institut f\"{u}r Strahlen- und Kernphysik of Bonn University, for warm hospitality where part of this work is being done; and thanks Bastian Kubis and Ulf-G. Mei{\ss}ner for helpful discussions. This work is supported in part by National Nature Science Foundations of China (NSFC) under Contract Nos. 10925522, 11021092; and by the Spanish Ministerio de Econom\'ia y Competitividad (MINECO) and the European Regional Development Fund (ERDF), under contracts FIS2014-51948-C2-1-P, FIS2014-51948-C2-2-P, FIS2017-84038-C2-1-P, FIS2017-84038-C2-2-P, SEV-2014-0398, and by Generalitat Valenciana under contract PROMETEOII/2014/0068.

\bibliographystyle{elsarticle-num}
\bibliography{piNletter}

\end{document}